SOFTWARE METAPAPER

# Fidimag – A Finite Difference Atomistic and Micromagnetic Simulation Package

Marc-Antonio Bisotti[1], David Cortés-Ortuño[1], Ryan Pepper[1], Weiwei Wang[2], Marijan Beg[3], Thomas Kluyver[1,3] and Hans Fangohr[1,3]

[1] Faculty of Engineering and the Environment, University of Southampton, Southampton, SO17 1BJ, UK
[2] Department of Physics, Ningbo University, Ningbo, 315211, CN
[3] European XFEL, 22869 Schenefeld, DE
Corresponding author: Marc-Antonio Bisotti (mb8g11@soton.ac.uk)

Fidimag is an open-source scientific code for the study of magnetic materials at the nano- or micro-scale using either atomistic or finite difference micromagnetic simulations, which are based on solving the Landau-Lifshitz-Gilbert equation. In addition, it implements simple procedures for calculating energy barriers in the magnetisation through variants of the nudged elastic band method. This computer software has been developed with the aim of creating a simple code structure that can be readily installed, tested, and extended. An agile development approach was adopted, with a strong emphasis on automated builds and tests, and reproducibility of results. The main code and interface to specify simulations are written in Python, which allows simple and readable simulation and analysis configuration scripts. Computationally costly calculations are written in C and exposed to the Python interface as Cython extensions. Docker containers are shipped for a convenient setup experience. The code is freely available on GitHub and includes documentation and examples in the form of Jupyter notebooks.

**Funding Statement:** We acknowledge financial support from EPSRC's Centre for Doctoral Training in Next Generation Computational Modelling, (EP/L015382/1), EPSRC's Doctoral Training Centre in Complex System Simulation (EP/G03690X/1), CONICYT Chilean scholarship programme Becas Chile (72140061), Horizon 2020 European Research Infrastructure project OpenDreamKit (676541), National Natural Science Foundation of China (11604169), and the Gordon and Betty Moore Foundation through Grant GBMF #4856, by the Alfred P. Sloan Foundation and by the Helmsley Trust.

## (1) Overview
### Introduction
The simulation of magnetic materials falls into several paradigms, depending on the length scales, materials and phenomena of interest. For many materials, the spin of atoms can be assumed to be localised around the atom, and a classical approximation can be made in which the atomic spin is treated as a point dipole – this is the classical Heisenberg model. Because the atomic crystal lattice of the material is considered, this discrete spin model is commonly referred to as atomistic [1], even though a semi-classical magnetic moment is assumed per lattice site. The continuum limit of this theory, known as micromagnetism, allows the computational treatment of much larger systems, though this excludes the study of materials which exhibit antiferromagnetism and ferrimagnetism.

Fidimag is a software library which allows researchers to model magnetic materials using both the classical Heisenberg and micromagnetic models. Users of the software provide the magnetic parameters of the material under study, the system geometry, and a set of initial conditions. The simulation can occur under different kind of dynamics, with Fidimag implementing the Landau-Lifshitz-Gilbert (LLG), the stochastic LLG (SLLG), and several spin-transfer torque variations of the LLG equation. From this initial setup, the user can then choose to evolve the system either through time, and hence study the magnetisation dynamics, or to "relax" the system to find its metastable energy states. In addition, the software implements the nudged elastic band method to find minimum energy paths and the size of energy barriers between different configuration states.

Fidimag has been used to obtain the results in several scientific publications [2, 3, 4, 5]. The supplementary data to ref. [4] demonstrates that with Docker, the process of correctly reproducing the results can be simplified to a single, reliable Makefile instruction [6].

### Standard Problem 4

The Micromagnetic Modeling Activity Group μmag has defined a number of Standard Problems for the validation and comparison of micromagnetic simulation software [7]. The definitions of the problems and solutions from different research groups have been published on its website, and aim to show differences in approach between different computational micromagnetic software. To this end, we have compared Fidimag to the published results, and these problems are also given as examples in the Fidimag documentation.

For illustration, we show here the solution of Standard Problem #4. In this example, the magnetisation reversal dynamics of a bar of Permalloy with the dimensions 500 × 125 × 3 nm are computed. First, a relaxed 's-state' is obtained in the absence of an applied magnetic field. It is plotted in **Figure 1a**. In a second simulation, a Zeeman field is applied that causes a reversal of the average magnetisation direction over the span of a few hundred picoseconds. **Figure 1b** shows the magnetisation configuration when the *z*-component of the spatially averaged magnetisation direction crosses 0. The system can be considered in equilibrium after a nanosecond and the corresponding magnetisation state is plotted in **Figure 1c**. The code used is printed in **Listing 1**.

```
import os
import matplotlib.pyplot as plt
import numpy as np
from fidimag.micro import Sim, UniformExchange, Demag, Zeeman
from fidimag.common import CuboidMesh
from fidimag.common.constant import mu_0

mesh = CuboidMesh(nx=200, ny=50, nz=1, dx=2.5, dy=2.5, dz=3, unit_length=1e-9)

A = 1.3e-11
Ms = 8.0e5
alpha = 0.02
gamma = 2.211e5
mT = 0.001 / mu_0

def compute_initial_magnetisation():
    sim = Sim(mesh, name='problem4_init')
    sim.driver.set_tols(rtol=1e-10, atol=1e-10)
    sim.driver.alpha = 0.5
    sim.driver.gamma = gamma
    sim.Ms = Ms
    sim.do_precession = False # saves time - not interested in dynamics here
    sim.set_m((1, 0.25, 0.1))
    sim.add(UniformExchange(A))
    sim.add(Demag())
    sim.relax(dt=1e-13, stopping_dmdt=0.01, max_steps=5000, save_m_steps=None, save_vtk_steps=None)
    return sim.spin

def compute_dynamics(initial_magnetisation):
    sim = Sim(mesh, name='problem4_dynamics')
    sim.set_m(initial_magnetisation)
    sim.driver.set_tols(rtol=1e-10, atol=1e-10)
    sim.driver.alpha = alpha
    sim.driver.gamma = gamma
    sim.Ms = Ms
    sim.add(UniformExchange(A))
    sim.add(Demag())
    sim.add(Zeeman([-24.6 * mT, 4.3 * mT, 0], name='H'), save_field=True)

    crossed_zero = False
    ts = np.linspace(0, 1e-9, 201)
    for t in ts:
        sim.driver.run_until(t)
        mx, my, mz = sim.compute_average()

        print("t = {:.3} ns\t mx = {:.3}".format(t*1e9, mx))
        if mx <= 0 and not crossed_zero:
            print("Crossed zero!")
            np.save("problem4_m_when_mz_0.npy", sim.spin)
            crossed_zero = True
    return sim.spin

def plot_quiver(m, filename):
    m.shape = (50, 200, 3)
    skip = 5
    m = m[1::skip, 1::skip]

    xyz = mesh.coordinates
    xyz.shape = (50, 200, 3)
    xyz = xyz[1::skip, 1::skip]
```

```
    plt.figure(figsize=(12, 3))
    plt.axes().set_aspect('equal')
    plt.quiver(xyz[:,:,0], xyz[:,:,1],
        m[:,:,0], m[:,:,1], m[:,:,1],
        pivot='mid', scale=45, cmap=plt.get_cmap('jet'), edgecolors='None')
    c = plt.colorbar()
    plt.xlabel("x (nm)")
    plt.ylabel("y (nm)")
    c.set_label("$m_\mathrm{y}$ (1)")
    plt.clim([0, 1])
    plt.xlim([0, 500])
    plt.ylim([0, 125])
    plt.savefig(filename)

if __name__ == '__main__':
    m0_file = "problem4_m0.npy"
    if not os.path.exists(m0_file):
        m0 = compute_initial_magnetisation()
        np.save(m0_file, m0)
    else:
        m0 = np.load(m0_file)

    mf_file = "problem4_mf.npy"
    if not os.path.exists(mf_file):
        mf = compute_dynamics(m0)
        np.save(mf_file, mf)
    else:
        mf = np.load(mf_file)

    plot_quiver(m0, "problem4_m0.pdf")
    plot_quiver(np.load("problem4_m_when_mz_0.npy"), "problem4_m_when_mz_0.pdf")
    plot_quiver(mf, "problem4_mf.pdf")
```

**Listing 1:** Solution to the standard problem #4. The simulation defined in the function `compute_initial_magnetisation` returns the relaxed magnetisation configuration in absence of an applied magnetic field. In the function `compute_dynamics` that makes up the second simulation, the magnetic field that causes the magnetisation reversal is added on line 42. The snapshots in Figure 1 were made using the code in the function `plot_quiver`.

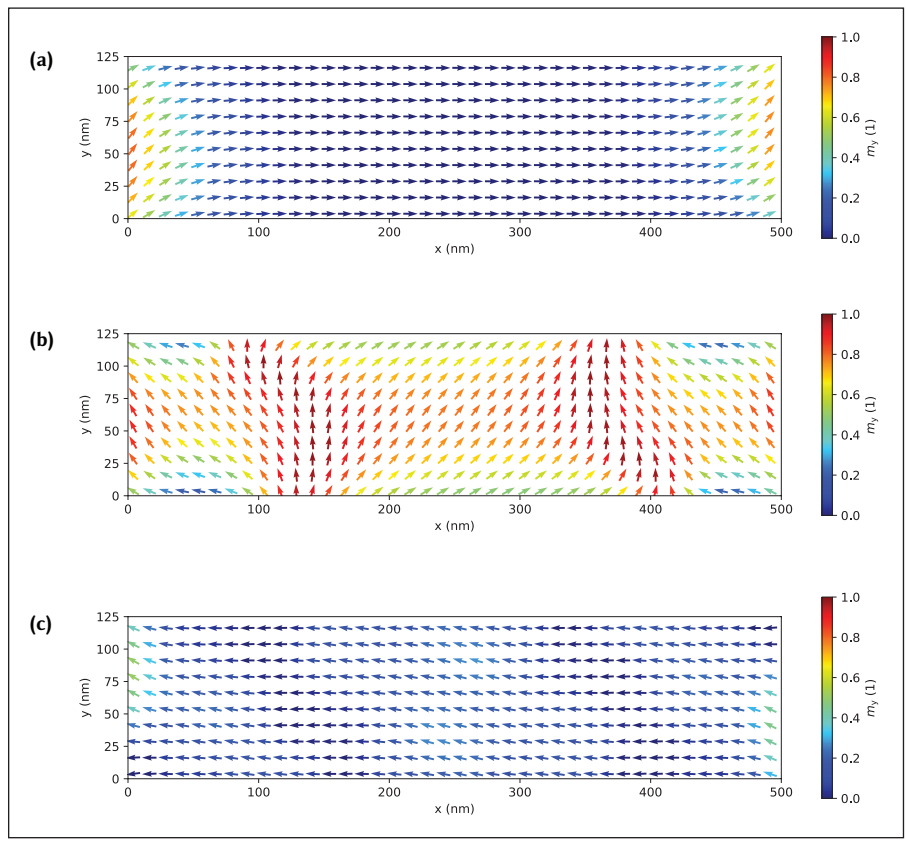

**Figure 1:** Snapshots of the unit magnetisation. The plots were created with the code in the function `plot_quiver` of listing 1. **(a)** The initial unit magnetisation of the bar of Permalloy described in standard problem #4. **(b)** The magnetisation conguration at the moment of the reversal of the average magnetisation. **(c)** The magnetisation after a nanosecond of simulation time has elapsed.

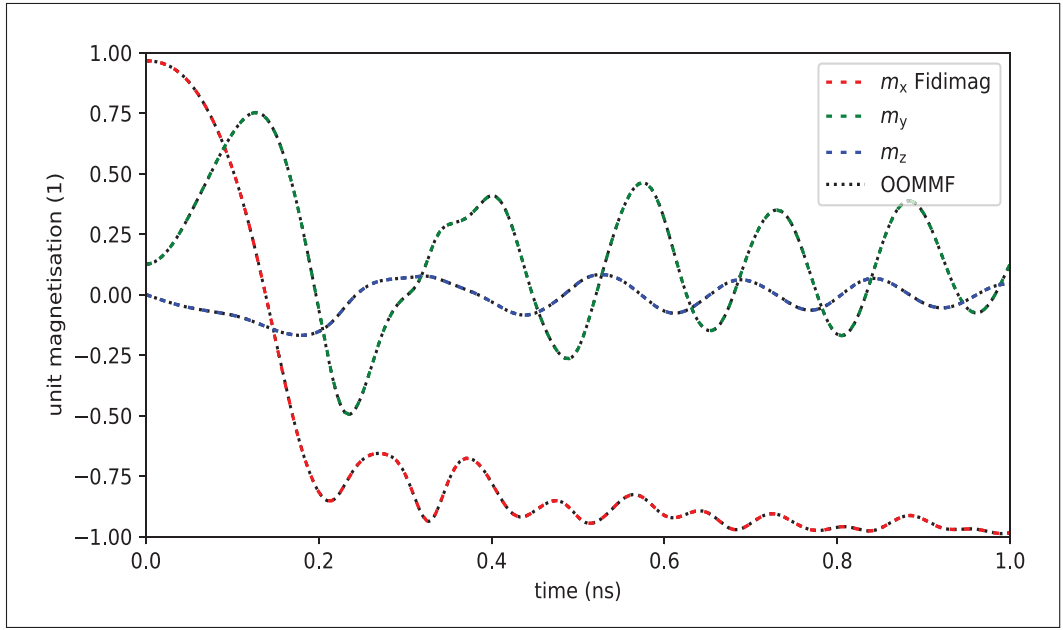

**Figure 2:** The components of the average magnetisation over time as computed with Fidimag and by OOMMF.

The evolution of the spatially averaged magnetisation is shown in **Figure 2** with a comparison with the same simulation run with OOMMF. This comparison is performed regularly as part of Fidimag's automated testing.

### *Computational simulations of magnetic materials*

At the atomic level, the magnetic effects in magnetic materials originate from two angular momentum terms from electrons: their orbital motion around the nucleus and, mainly, from a quantum property of electrons called spin.[1] Within a semi-classical model, the sum of their contributions can be described by a three dimensional vector $\mathbf{S}_i = \mu_i \mathbf{s}_i$ representing a magnetic moment (a magnetic dipole) $\mu_i$ for every atom $i$, with a specific direction $\mathbf{s}_i$. This idea originates from Heisenberg's model Hamiltonian to describe magnetic interactions [8, 9]. In magnetic solids, atoms arrange in a lattice with a specific periodic crystal geometry. For example, a cubic lattice is made of repeated cells where atoms and their corresponding magnetic moments lie at the corners of this cubic cell.

An atom is at the scale of a few Angstrom, thus in large systems, which are at the nano-scale, it would be necessary to describe the system using thousands of spins, which is computationally expensive. Therefore, it is possible to approximate the material as a continuum, where the field from discrete magnetic moments turns into a continuum field called magnetisation that depends on the space coordinates $\mathbf{r}$, assuming that neighbouring spins do not change drastically in direction [8, 9]. This means the limit from the discrete atoms into the space dependent field $\mathbf{s}_i \rightarrow \mathbf{M}(\mathbf{r})/M_s = \mathbf{m}(\mathbf{r})$. The magnitude of the field is measured through the saturation magnetisation $M_s$, which is defined as a magnetic moment per unit volume, i.e. $M_s = \mu/\Delta V$, with $\Delta V$ as the volume of a unit cell of the discrete crystal lattice. This theory of magnetism in the continuum limit is known as micromagnetics. Numerically, the magnetisation field can be discretised into a mesh of magnetic moments $\mathbf{M}_i$.

To describe a magnetic system, we firstly specify its geometry. If we simulate this system using a discrete spin model, we have to generate a lattice of magnetic moments with a specific arrangement of atoms (according to the crystal symmetry) such that they describe this geometry. In the case of micromagnetics we use the finite differences numerical technique, thus the sample is discretised into cuboids and each one of them represents a volume with uniform magnetisation inside that volume.

### *Dynamics*

Magnetic phenomena emerge from the interactions between magnetic moments and their interaction with external and internal magnetic fields, the later including dipolar interactions and anisotropic interactions, among others. These interactions depend on the material and specify the total energy of the system. In general, the spins perform a precessional motion following directions set by the magnetic interaction. The dynamics of the magnetic moments is given by the Landau-Lifshitz-Gilbert (LLG) equation [8, 9]. Within the discrete spin model this dynamical equation for a single spin **s**, reads:

$$\frac{\partial \mathbf{s}}{\partial t} = -\gamma \mathbf{s} \times \mathbf{H}_{\text{eff}} + \frac{\alpha_G \gamma}{\mu} \mathbf{s} \times \mathbf{s} \times \mathbf{H}_{\text{eff}}, \quad (1)$$

at zero temperature, where $\mathbf{H}_{\text{eff}}$ is an effective field that contains the sum of every magnetic interaction present in the system, $\gamma$ is the Gilbert gyromagnetic ratio constant, which sets the time scale of the spin motion, and $0 \leq \alpha_G \leq 1$ is the Gilbert damping. The first term in equation 1 describes a precessional motion of spins around the effective field and the second term is a dissipative term that make spins follow the effective field direction.

In micromagnetics this equation has the same structure:

$$\frac{\partial \mathbf{M}}{\partial t} = -\gamma \mathbf{M} \times \mathbf{H}_{\text{eff}} + \frac{\alpha_G \gamma}{M_s} \mathbf{M} \times \mathbf{M} \times \mathbf{H}_{\text{eff}}. \quad (2)$$

Fidimag can solve the non-linear differential equations 1 and 2, depending on the specified theoretical model. Accordingly, it is necessary to specify an initial magnetic configuration for the spin directions, and the magnetic interactions involved in the system. Currently, Fidimag has the following interactions implemented in the code,

- **Exchange** Favours the parallel or anti-parallel alignment of neighbouring spins. The exchange can be computed using nearest neighbour approximations in the classical Heisenberg case. The continuum micromagnetic model only allows parallel alignment.
- **Dipolar interactions** Product of the interaction of a spin with the field generated from the other spins in the system.
- **Dzyaloshinskii-Moriya** Favours a canted alignment between neighbouring spins.
- **Anisotropy** Sets directions along which the system energy decreases when spins align towards them.
- **Zeeman** Product of the interaction of spins with an external magnetic field.

The underlying equations that were used to implement these interactions can be found in Ref. [10].

One method for finding energy minima in the system's energy landscape, which is implemented in the code, is to evolve the system with the LLG equation, since the spins will precess until a stable configuration is attained, *i.e.* a local or global minimum of energy, that depends on the initial magnetic configuration. We refer to this process as relaxation.

Dynamical effects such as the generation of spin waves, ferromagnetic resonance, domain wall motion, among other multiple magnetism related phenomena, are also given by the evolution of spins with the LLG equation. Variations of this equation are obtained when applying electric currents or temperature, which can also be specified in the code. Fidimag supports the inclusion of stochastic terms and spin-polarised currents in the equation of motion. The terms and their corresponding equations are shown in Ref. [11].

At zero temperature the length of the magnetic moments and the magnetisation vector is fixed. This condition must be specified in the equation of motion of spins explained in the last section, which sets a constraint to the spin length. Multiple magnetic phenomena can be explained at a zero-temperature formalism thus the majority of Fidimag's equations are implemented in this regime.

Although this constraint is implicit in the LLG equation, numerically the spin length varies when integrated. To address this problem, a correction term is added to the right hand side of equations 1 and 2 in Fidimag (equation 4 in Ref. [12]) [13]. It is proportional to:

$$\sqrt{\left(\frac{\partial \mathbf{s}}{\partial t}\right)^2}\left(1-\mathbf{s}^2\right)\mathbf{s}. \quad (3)$$

and similar in the micromagnetic model, by setting $\mathbf{s} \rightarrow \mathbf{m}$. This term makes the spin length increase when it is less than the unit and makes it decrease otherwise. For time integration, Fidimag can use the CVODE solver of the SUNDIALS suite [14] (wrapped via Cython) or the Fortran codes bundled in Scipy's [15] `integrate.ode`. Fidimag also implements Heun's method and the classic fourth-order Runge-Kutta method. Because the time integrators will need a varying amount of evaluations of the right hand side function, *i.e.* the equation of motion chosen in the last paragraph, it makes sense to partly relinquish runtime control from the user to the time integration. A so-called driver is in charge of coordinating time integration and user needs. Time integration runs uninterrupted for a specified amount of (simulated) time, after which arbitrary code can be executed.

### *Nudged Elastic Band Method*

Finding the lowest energy cost to drive a magnetic system from one equilibrium state towards another, also known as energy barrier, has become a relevant problem for the analysis of the stability of magnetic structures. An energy barrier is then associated to the transition path, between two states, that requires the least energy. This is important, for example, for the potential application of a magnetic structure in a technological device since an energy barrier can be used to estimate the lifetime of the structure against energy fluctuations from excitations such as thermal noise, present at finite temperatures. The nudged elastic band method (NEBM) is a technique for the calculation of minimum energy paths, and hence energy barriers. It was first used in chemistry to study molecular transitions [16]. It is based on fixing two equilibrium states, which can be local or global minima, and making copies, called images, of the system in different configurations between these extrema. This sets a *band* of images and each one of the images will have a different energy in the energy landscape associated to the system. The algorithm will iteratively find a path in the energy landscape that decreases the energy cost of the *band*, trying to keep the images equally spaced in the energy landscape using an inter-image spring force, to avoid images move towards the minima at the extremes of the band. After relaxation, the minimum energy path will cross one or more maxima of energy, which set the energy barrier magnitude. An optimisation of the original algorithm has recently been published by Bessarab et al. [17], where geodesic distances are defined in the energy landscape. We have implemented the NEBM in Fidimag, which can be used both within micromagnetics and a discrete spin model. The optimised version, called Geodesic NEBM is the one that performs more efficiently, and combines Cartesian coordinates for the description of the spins and geodesic distances in the energy landscape. Original versions of the algorithm, which only use Cartesian or spherical coordinates, can present convergence issues for systems with more complex magnetic configurations, such as vortices. To run a NEBM simulation, it is necessary to specify two equilibrium states, the number of images in between, and an initial state for the internal images. This initial configuration can be set by either manually creating a series of images in different magnetic states or by using a linear interpolation for the spin directions, which is implemented in the code to generate the configurations automatically.

A detailed review of the method can be found in Ref. [17]. A study of the thermal stability of skyrmions, which are vortex-like magnetic configurations, that uses the NEBM implemented in Fidimag, can be found in [4]. In addition, we tested our code with the skyrmions example from [17], making a repository with the simulation details in [18].

**Implementation and architecture**

Fidimag provides both micromagnetic and atomistic simulation capabilities and is the only software that allows switching between the two modelling paradigms. The most notable micromagnetic code is OOMMF, which had its alpha release in 1998. OOOMMF is written in C++ and Tcl/Tk. Fidimag is mostly implemented in the Python programming language, with performance critical parts realised in C and linked in via Cython. The code base consists of roughly 5000 lines of Python code and 4000 lines of C code and cython extensions.[2] The tests and examples add another 5500 lines of Python. As Python is well established in the scientific community, it presents a lower barrier to entry for programmers [19]. Further, Python has a reputation for being easy to learn for beginners [20, 21]. This has informed the decision to build Fidimag as a software library instead of a GUI-driven program. Indeed, Fidimag respects best practices for Python's module and namespacing system.

This library model of execution, where Fidimag is imported into the namespace of a Python program empowers the user to deal with the complexity of the batch processing of simulations in a more easily tested and reproducible way compared to ad-hoc shell commands and specialised batch modes in software with graphical user interfaces [22].

The standard problem #3 especially examplifies how naturally Fidimag allows higher order logic thanks to these choices. In it, two possible magnetic configurations in cubes of increasing sizes are to be studied to determine when they are equally energetically favourable. In **Listing 2** the logic of setting up and running the simulations, as well as comparing the total energies of the magnetisation in the two possible configurations has been abstracted into a function called `energy_difference`.

Finding the cube size of equal energies called `single_domain_limit` then involves nothing more than another function call, this time to a bisection method called `bisect` provided by SciPy [15].

A similar approach had been used in Nmag, where it has proven to be successful [13]. Unlike nmag however, Fidimag doesn't require bundling the software with a modified Python binary. The approach is now seen in other micromagnetic packages, for example micromagnum [23]. The Jupyter/OOMMF project follows a hybrid approach and harnesses the proven capabilities of the OOMMF binary with a Python interface optimised for Jupyter notebooks [24].

Vampire is a performant atomistic code that defines its own declarative syntax [1]. It comes in two versions, either a binary with an installation script or the code which needs to be compiled from source. In contrast, while the traditional installation methods are available, installing Fidimag with Docker is as easy as `docker pull fidimag/notebook`.

Best practices recommend using version control for any kind of computational endeavours [25]. The code is stored using the distributed version control system git and new features are developed in branches. Automated testing and builds of the software were priorities from the start. Because there are no command line parameters to record or options set in a GUI program Fidimag offers a one-to-one mapping of the simulation code to its result. This is a boon for the goal of reproducibility that can't be overstated. For example, Fidimag plays well with Sumatra [26] which allows for the automatic tracking of the computations run. Jupyter Notebooks [27] are used for interactive tutorials and included in the extensive documentation [28]. There are tests that ensure that these notebooks are executing consistently, without errors, and that the execution of the stored inputs match the stored outputs. For this, a py.test plugin called nbval was developed and made available to the open source community [29]. Fidimag has an automated testing procedure that runs on a public continuous integration system [30]. Fidimag's simple installation procedure via Docker, the extensive documentation including interactive tutorials stored in Jupyter notebooks and the accessible interface have made it not only a useful tool for research as stated before, but also for teaching [31].

```
"""
Solution to μmag standard problem #3 with fidimag.
   http://www.ctcms.nist.gov/~rdm/mumag.org.html

"""
from .cube_sim import energy_difference
from scipy.optimize import bisect

single_domain_limit = bisect(energy_difference, 8, 8.5, xtol=0.1)
print("L = {:.2} nm.".format(str(single_domain_limit)))
```

**Listing 2:** Solution to the standard problem #3. The code that computes the energy difference between the two possible magnetic configurations has been abstracted into the function `energy_difference` in the module `cube_sim` that is imported in line 6. The function `energy_difference` can then be handed off to a bisect method provided by SciPy to find the cube size for which both configurations have the same energy.

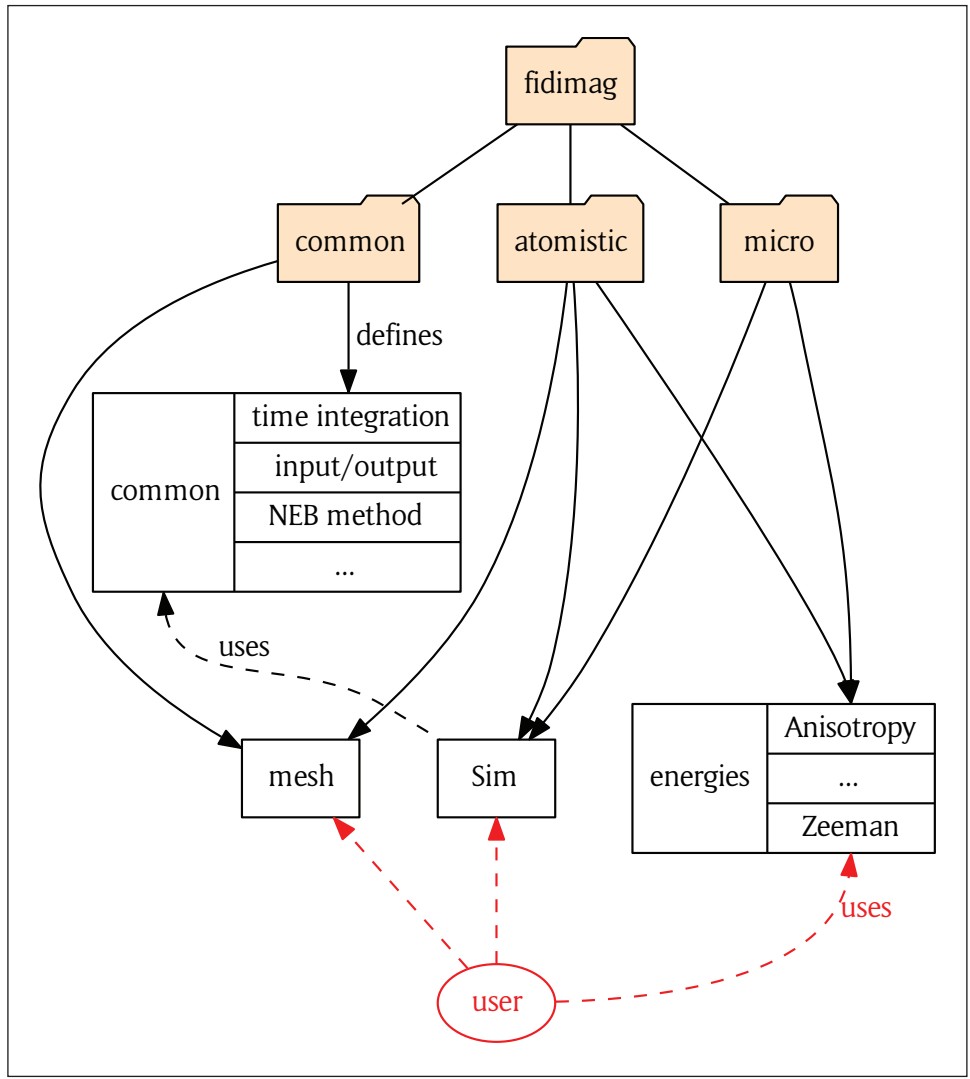

**Figure 3:** Architecture of Fidimag as reflected by the directory structure and Python modules. Code useful to micromagnetic and atomistic simulations was extracted into a `common` namespace. The red arrows point to user-facing parts of the system.

The source code is split into three major sections – 'atomistic', 'micro' and 'common' which contain code specifically for classical Heisenberg simulations, micromagnetic simulations, and for both respectively, as can be seen in **Figure 3**.

There are many common components which can be shared between both types of simulation. The primary data in both atomistic and micromagnetics is stored in vector fields, and the mechanisms for passing this data through the simulation are kept in a common base 'Sim' class. This allows, for example, the saving of the magnetisation progressive steps in the simulations to be handled identically for both types of simulation. Fidimag comes with visualisation and data saving utilities out of the box.

*Mesh and lattices*
As we mentioned earlier, we define a magnetic system by setting its geometry. How the geometry is approximated depends on the chosen model.

Within the continuum approximation, Fidimag uses finite differences which means dividing the sample into a mesh of cubes, as shown in **Figure 4a**, where each cuboid centre represents the position of a magnetic moment vector. Accordingly, derivatives for the calculation of magnetic interactions and energies are discretised using differences between neighbouring mesh sites. The cuboid mesh is coded in the `CuboidMesh` class located in Fidimag's `common` directory. Distance between the centres and the number of cuboids in the three spatial directions are specified by the user. The indexes of every mesh site are labelled following the $x \rightarrow y \rightarrow z$ direction, as shown in **Figure 4a**. To keep track of the neighbouring sites of every mesh site, the mesh class has a `neighbours` method defined as an $N \times 6$ array, where $N$ is the total number of cuboids, that stores the indexes of nearest neighbours, with the value –1 for non-material sites. Correspondingly, every row represents the 6 nearest neighbours in the $(-x, +x, -y, +y, -z, +z)$ order. For instance, for the mesh of **Figure 4a** and site 0 the `neighbours` array is [–1,

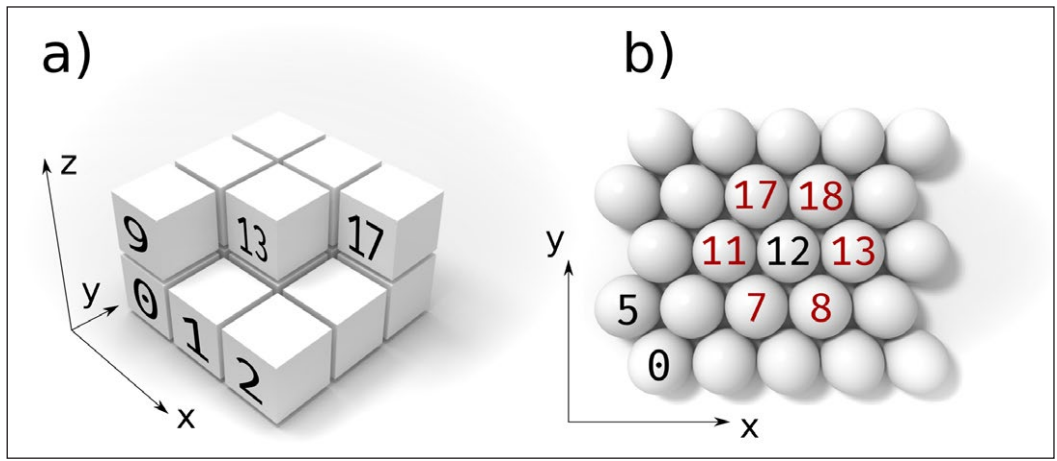

**Figure 4:** Cuboid mesh and hexagonal lattice labelled according to the index of the sites.

1, –1, 3, –1, 9]. This helps to easily implement interactions in Fidimag's C functions.

For the calculation of the demagnetising field, Fidimag uses a fast-Fourier-transform method that requires a uniform grid, hence for complex geometries, such as curved samples, Fidimag still defines a cuboid mesh but boundaries or specific mesh sites without material are approximated by setting the magnetisation as zero, $M_s = 0$. Thus to obtain a better approximation it is necessary to define a large number of cuboids and decrease the distance between their centres. In order to set a null magnetisation, the user can specify a Python function that has as an input the position vector and that returns $M_s$ according to the sample geometry [13].

In **Listing 3** we set a 2 nm diameter sphere using a 6 × 6 × 6 cuboid mesh with elements of 1nm × 1nm × 1nm size. To simplify the dimensions scaling we set a unit length in the mesh definition.

```
import fidimag
import numpy as np

mesh = fidimag.common.CuboidMesh(nx=6, ny=6, nz=6,
                                  dx=1, dy=1, dz=1,
                                  unit_length=1e-9)
sim = fidimag.micro.Sim(mesh)
Ms = 1e6 # A / m

def sphere(r):
    x, y, z = r
    if x ** 2 + y ** 2 <= 2 ** 2:
        return Ms
    else:
        return 0

sim.set_Ms(sphere)
```

**Listing 3:** Definition of a 2 nm wide sphere geometry using micromagnetics.

In the case of discrete spin simulations, atoms order in a lattice according to different crystallographic arrangements. The most simple ordering is the *simple cubic* lattice (SC), where atoms centres lie at the centre of cubes separated by the same distance in the three spatial dimensions. Since this follows the same principle as the finite differences cuboid mesh, Fidimag uses the same `CuboidMesh` class as the one for micromagnetic simulations, thus it is defined in Fidimag's `common` directory. Another atomistic lattice implemented in Fidimag is a two dimensional hexagonal lattice that can be aligned along the x or y axis. The former is shown in **Figure 4b**. This lattice is defined in the `HexagonalMesh` class in Fidimag's `atomistic` directory. As in the `CuboidMesh`, a `neighbours` array keeps track of the neighbours indexes, which are defined in the $x \to y$ order. In **Figure 4b** we highlight, for example, the 6 neighbours of the lattice site with index 12.

Using as a base the class structure of the meshes currently implemented in Fidimag, it is possible to define and implement other crystal lattices such as body-centered-cubic or face-centered-cubic arrangements.

Besides scalars, most of the physical quantities involved in micromagnetic computations are fields and vector fields. In Fidimag, they are represented with numpy arrays [32]. There are utilities to instantiate them from point-wise expressions. These fields are the objects the physical interactions in the simulations deal with. A quantity central to the solution of the initial value problem given by the micromagnetic approach is the effective field, which is itself a sum of smaller fields. Each of the smaller fields is the result of a separate computation which represents the influence of an interaction present in the system, like the exchange energy or the Dzyaloshinski-Moriya interaction. For the dipolar field, we accelerate the calculation using a convolution done in Fourier space [33, 34, 35], accelerated using the library FFTW. It is in the computation of the interactions where the differences between the micromagnetic and the atomistic approach become the most apparent. Consequently, some energy terms are represented twice in the codebase – once for each model. With the effective field computed for a given state of the magnetisation the next step is solving the equation of motion for the system, which is the Landau-Lifshitz-Gilbert equation or one of its many variants.

### Quality control

Testing small chunks of code and preferably isolated pieces of the system is unit testing. In Fidimag, most of the unit tests compute and check the physical quantities involved in succesfully running a simulation. Besides giving helpful feedback during development, the unit tests

```
def test_save_scalar_field_hexagonal_mesh(tmpdir):
    mesh = HexagonalMesh(radius=1, nx=3, ny=3)
    s = scalar_field(mesh, lambda r: r[0] + r[1])
    vtk = VTK(mesh, directory=str(tmpdir), filename="scalar_hexagonal")
    vtk.save_scalar(s, name="s")
    assert same_as_ref(vtk.write_file(), REF_DIR)
```

**Listing 4:** A unit test in which a scalar field `s` is created on a hexagonal mesh, saved to a VTK file and compared to a reference file.

were a factor in increasing confidence to edit, enhance or refactor the code, resulting in the improved decoupling and maintainability of it. The coverage of Fidimag's code base with unit tests is monitored using codecov [36]. At the time of writing, it reports 73% coverage.

Functional testing then examines a slice of the system. For example, one of the test cases covering the saving of scalar fields on hexagonal meshes to a VTK file in Fidimag is shown in **Listing 4**.

Finally, system tests use Fidimag as a black box and compare simulation output to known good values. These values are obtained from problems that have analytical solutions like the macrospin or domain walls. As we saw earlier, the $\mu$mag standard problems are another good source of testing data. So are other finite difference codes, like the aforementioned OOMMF.

Fidimag is also tested against the output of earlier versions of itself by storing some simulation results in the repository. This is a form of regression testing, and since no external software or potentially long-running simulations are involved the fastest way to check if a changeset has affected the computational parts in a tangible way. The design choices discussed in the introduction and the focus on testing continue to provide tangible benefits to the collaborators working on the software, as well as the software itself, as shown above. They also increase the confidence in Fidimag›s results since unlike users, automated tests don›t distinguish between new and old code and test potentially seldom-used parts of the software just as often as the commonly used functions.

For maximum value, the tests need to be run often and ideally without manual intervention. This is why a test cycle is triggered on the continuous integration platform Travis CI [30] on every push to the Github repository. First, Fidimag's C code and cython extensions are compiled. Then the tests are run. Finally the Jupyter notebooks are run and tested and the documentation is built.

To quickly check if Fidimag is installed and working on a machine, a user can launch an interactive Python shell and execute `import fidimag`. This command should not output any text. Afterwards, the user may chose to follow along with the tutorial called *A Basic Simulation* [37] or launch the provided examples. If Fidimag was installed from source, `make test-basic` will run a selection of tests that completes in under a minute, which is also helpful for smoke testing purposes during development on Fidimag.

## (2) Availability
**Operating system**
GNU/Linux, Mac OS X and on any platform supported by Docker, like Azure and AWS.

**Programming language**
Python version 3, C, cython.

**Additional system requirements**
E.g. memory, disk space, processor, input devices, output devices.

**Dependencies**
E.g. libraries, frameworks, incl. minimum version compatibility.

1. Numpy ≥ 1.10
2. SciPy ≥ 1.0.0
3. SUNDIALS ≥ 2.7.0
4. FFTW ≥ 3.3.4

Fidimag builds using the OpenMP versions of these libraries for multiprocessing purposes.

**List of contributors**
Please list anyone who helped to create the software (who may also not be an author of this paper), including their roles and affiliations.

1. Bisotti, Marc-Antonio
2. Wang, Weiwei
3. Cortés-Ortuño, David
4. Fangohr, Hans
5. Pepper, Ryan
6. Kluyver, Thomas
7. Vousden, Mark
8. Beg, Marijan

**Software location**
*Archive*
(e.g. institutional repository, general repository) (required please see instructions on journal website for depositing archive copy of software in a suitable repository)
  **Name:** Fidimag
  **Persistent identifier:** https://zenodo.org/record/841113
  **Licence:** BSD
  **Publisher:** Hans Fangohr
  **Version published:** 2.5
  **Date published:** 10/08/17

*Code repository*
(e.g. SourceForge, GitHub etc.) (required)
  **Name:** Fidimag
  **Persistent identifier:** https://github.com/computationalmodelling/fidimag
  **Licence:** BSD
  **Date published:** August 8, 2018

*Emulation environment*
(if appropriate)
   *Name:* The name of the archive.
   *Persistent identifier:* e.g. DOI, handle, PURL, etc.
   *Licence:* Open license under which the software is licensed.
   *Date published:* dd/mm/yy

**Language**
Language of repository, software and supporting files. English.

## (3) Reuse potential

The simulations are written in Python and use Fidimag as a library to be imported. Fidimag has been used to gather results for a number of scientific publications [2, 3, 4, 5] in the field of micromagnetics. It can be extended to account for other energy terms, to support more time integration methods or other variants of the equation of motion, as well as other drivers. Users can get in touch with the development team on Fidimag's issue page on GitHub [38].

**Notes**
[1] For this reason when we mention spins we strictly refer to the total angular momentum or magnetic moments.
[2] Numbers obtained with cloc version 1.60.

**Competing Interests**
The authors have no competing interests to declare.